# Elastic Coulomb-levitation: why is ice so slippery?

Chang Q Sun

ecqsun@ntu.edu.sg

Nanyang Technological University, Singapore

**The elastic, less dense, polarized, and thermally stable supersolid skin lubricates ice. Molecular undercoordination shortens the H-O bond and lengthens the O:H nonbond through O-O repulsion, which is associated with low-frequency and high-magnitude of O:H vibration and a dual O-O polarization. The softer O:H springs attached with stronger molecular dipoles provide forces levitating objects sliding on ice, like Maglev or Hovercraft.**

## Contents

# 1     Anomaly: Ice is most slippery of ever known

Ice is most slippery of ever known at even temperatures below the limit of pressure melting, -22 °C. Ice and snow abound when winter weather hits, all sorts of surfaces can get slick and slippery. Skating on ice is a jealous entertainment, see Figure 1. However, if you are a driver, this is quite troublesome. Ice and snow can make driving treacherous. Slipperiness of snow and ice forms the platform of Winter Olympic Games but it is one of the unanswered puzzles about ice.

The general consensus since 1850 [1] is that there is a thin liquid or quasiliquid layer covering ice to act as a lubricant. Existing mechanisms of: 1) pressure melting [2, 3], 2) friction heating [4], or 3) quasiliquid skin due to undercoordinated molecules [5] could hardly explain how this liquid layer forms.

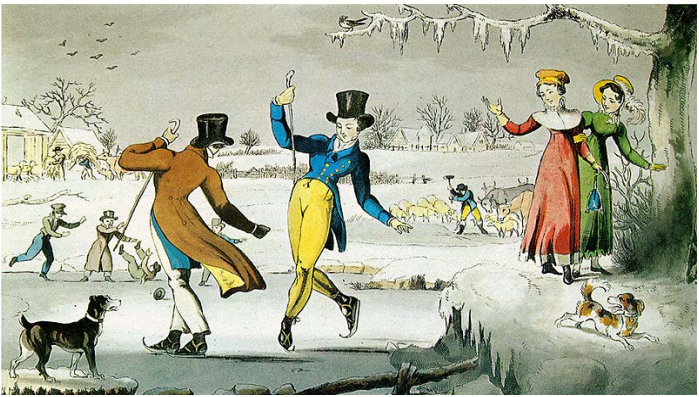

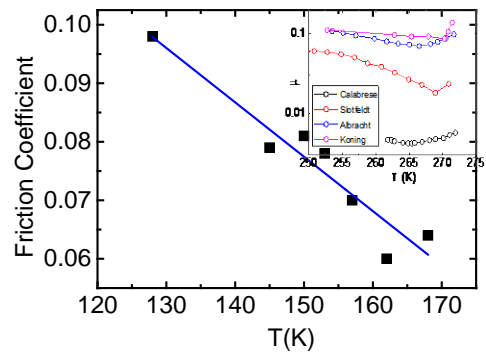

Figure 1. Is ice covered with a quasiliquid or a supersolid skin? (a) An early 1820's print for the ice-skating scene (published in London). The skate blades are strapped on to the shoes rather than being an integral part of the modern skates. (Credit: W. Belen, Free Wikipedia.) (b) The friction coefficient of steel-pin on ice-disc under $10^{-10}$ Pa vacuum condition shows linear temperature dependence in the regime of solid bulk phase [6]. Inset shows friction trends in the bulk quasi-solid phase regime [7] under different conditions [8]. (Reprinted with permission from [6].)

## 2    Reasons: Elastic Coulomb-levitation

Instead of a quasiliquid layer, ice is covered by elastic, polarized, less dense and thermally more stable supersolid skin [9-11], as illustrated in Figure 2:

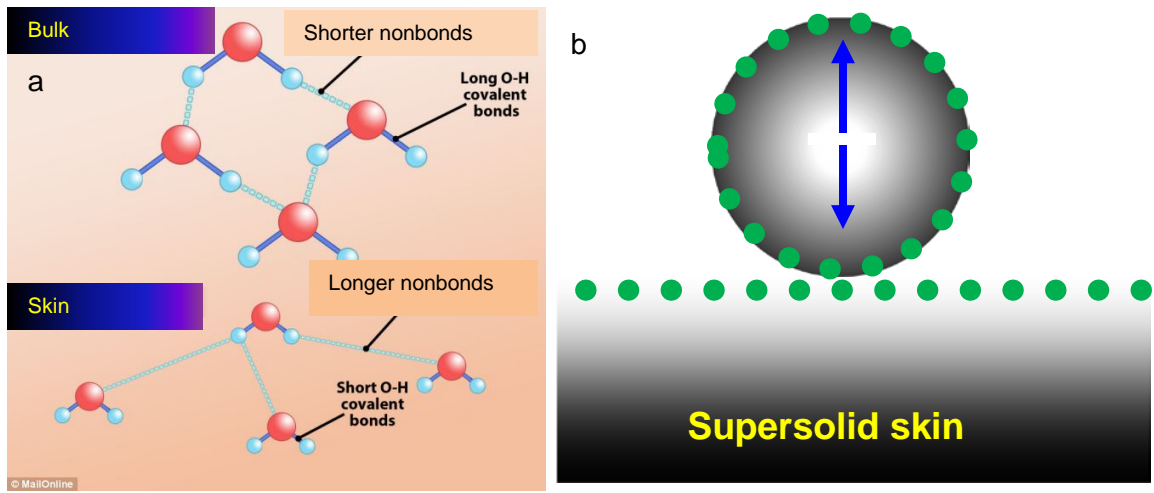

Figure 2. Elastic Coulomb-levitation makes ice's supersolid-skin slippery. (a) Molecular undercoordination reduces their sizes but enlarges their separations of $H_2O$ molecules, which softens the O:H nonbond by lowering the frequency but enhancing the amplitude of O:H vibration; molecular undercoordination does oppositely to the H-O bond [10]. The longer and softer O:H springs attached with dipoles due to dual polarization not only levitate the object on it but also recover readily from deformation, which make the supersolid ice skin elastic and slippery. The sliding object is also negatively charged eventually despite dielectric substance. Arrows denote the force acting on the load: $F_N + F_C - mg = 0$, with $F_N$, $F_C$, and $mg$ being the normal force, the Coulomb levitation, and the weight of the object, respectively.

1) Molecular undercoordination shortens and stiffens the H-O bond, and meanwhile, lengthens and softens the O:H nonbond with a dual polarization of electrons on oxygen ions (H-O contraction polarizes the lone pair electrons in the first round and that then enhances O-O repulsion in the second).

2) H-O bond stiffening raises the melting point from 273 to 310 K and the H-O phonon frequency from 3200 to 3450 cm$^{-1}$; O-O elongation lowers the local mass density from 1.0 to 0.75 g·cm$^{-3}$.

3) The O:H nonbond softening and the O-O dual polarization enhance the viscoelasticity and hydrophobicity of the skin.

4) Coulomb repulsion between the locally pinned dipoles attached to O:H springs lowers the friction at the interface contact, making ice slippery, which is the same in principle to Maglev train and Hovercraft, see Figure 3.

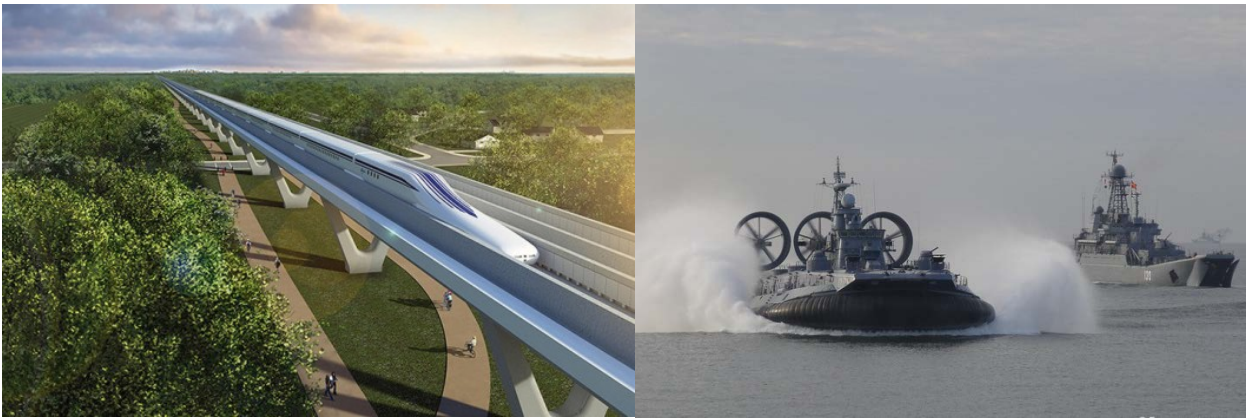

Figure 3. Maglev train and Hovercraft move faster because of the lowed friction by introducing magnetic repulsion for the former and flowing air the latter at the contacting interfaces. (Free Wikipedia)

3 History background

3.1 Ice friction

The first report of sliding on ice comes from Scandinavia Mountains, the source for repeated glaciation that covered much of eastern, central and western Europe with a particular emphasis on Denmark, Norway, Sweden and Finland, around 7000 B.C. Rock carvings illustrate the use of a sledge for the transport of heavy goods. The next interesting historic record dates back to 2400 B.C. Egyptian carvings showed that a lubricant, possibly water, was poured in front of a sledge to facilitate sliding [8].

Chinese was the first to have moved a chunk of mountain along a road of man-made ice. In the 15th century [12], Chinese architectures transported large rocks weighing hundreds of tons to the site from 70 km away by using an artificial ice path to build the Forbidden City, an imperial palace, consists of about a thousand buildings, see Figure 4 for a typical building. The artificial path was made by pouring water from wells dig aside the path of certain separation each along the way. Chinese-wheeled carriages would not have been able to transport such heavy blocks, even with the technology of the late 1500s. Another option would be to use wooden rollers, but that would require creating a smooth surface on tricky, winding roads.

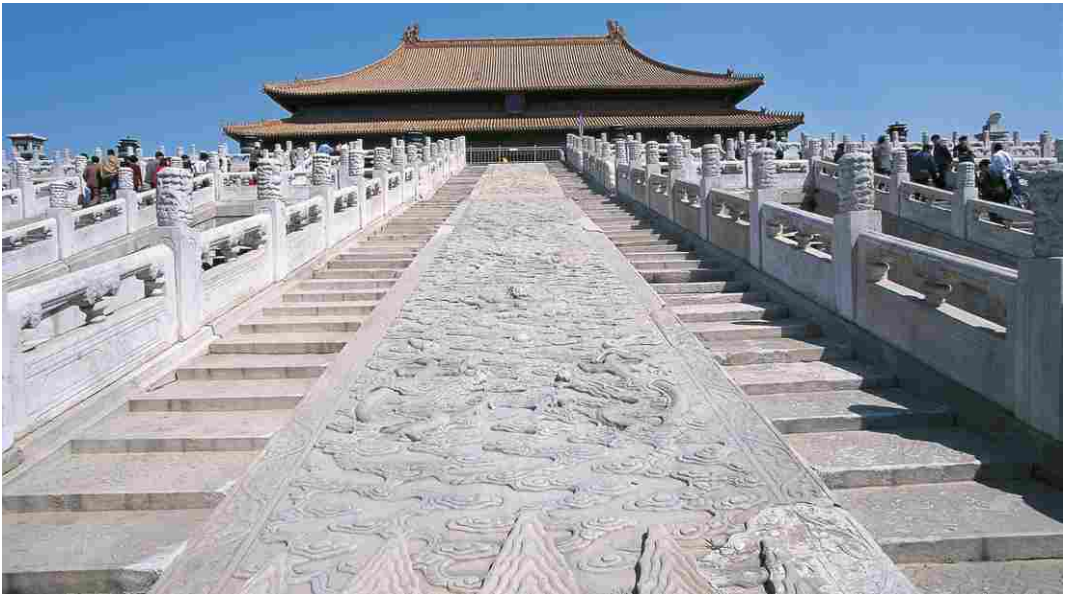

Figure 4. The Large Stone Carving is the heaviest stone in the Forbidden City in Beijing. It weighed more than 300 tons when it was first transported to the site between 1407 and 1420. (*DEA/ W. Buss/De Agostini/Getty Images*)

In 1785, Charles Augustin Coulomb, see Figure 5, investigated five main factors for frictional resistance. He studied the nature of materials in contact and surface coatings, the extent of the surface area, the normal pressure, the length of time that surfaces stay in contact, and the frictional behavior under vacuum as well as under varying ambient conditions namely temperature and humidity [8] . Besides, surface roughness, surface structure, wettability, sliding velocity, and thermal conductivity affect the friction behavior of ice. Applying electric field cross the contacting interface can also affect the coefficient of friction [13].

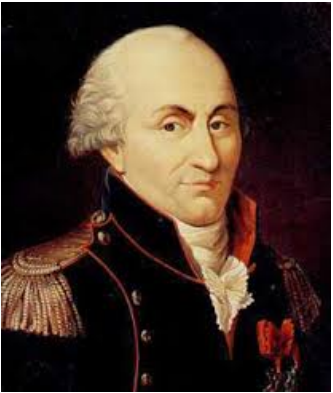

Figure 5. Charles-Augustin de Coulomb (14 June 1736 – 23 August 1806), a French physicist, was best known for developing Coulomb's law, the definition of the electrostatic force of attraction and repulsion, and on friction. (Free Wikipedia)

At the atomic scale, researchers [14] found that atomic–lattice vibration and the electronic-charge play significant roles in friction. When atoms close one surface are set into motion of atoms in the opposite surface creates waves in terms of phonons. The amount of mechanical energy transformed into phonons depends on the sliding substances. Solids are much like musical instruments in that they can vibrate only at certain distinct frequencies, so the amount of mechanical energy consumed will depend on the frequencies actually excited. If the "plucking" action of the atoms in the opposite surface resonates with one of the frequencies of the other, then friction arises. But if it is not resonant with any of the other surface's own frequencies, then sound waves are effectively generated. On the other hand, the smaller the resulting amplitude of vibration is, the greater the friction will be from the "rubbing" action of the film sliding about on the substrate. For insulating surfaces, friction arises from the attraction of unlike charges attached to the surfaces, like a balloon being rubbed on hair and left to cling to a wall. In 1989, Jacqueline Krim and coworkers found the friction coefficient Krypton films on crystalline gold surfaces is lower when dry. Adding a liquid film raises the coefficient by five times, instead [14].

Understanding mechanisms of friction on ice is particularly important in a broad field of applications, such as motorized vehicle traffic in winter road conditions, glacial movements, cargo transportation through northern sea ways, design of offshore structures and ice breakers, and ice sports. High friction on ice is desired for motorized vehicle traffic in winter road conditions and the grip of shoe soles on ice to avoid accidents. However, in the field of cargo transportation through northern sea ways and the design of offshore structures low friction materials are desired to limit maintenance and operation costs, e.g., 70% of the power of an ice breaker ship is consumed to overcome ice friction.

Sukhorukov and Loset [15] examined the effects of the sliding velocity (6-10$^5$ mm/s), air temperatures (-2 to -20 °C), normal load (300-2000 N), presence of sea water in the interface, and ice grain orientation with respect to the sliding direction on the friction coefficient of sea ice on itself. The kinetic friction coefficient of sea ice on sea ice varies from 0.05 (at – 20 °C) to 0.5 (at – 2 °C), regardless the presence of sea water in the sliding interface. The friction coefficient is independent of the velocity when sliding occurs between natural ice surfaces. As the contacting surfaces became smoother, the kinetic friction coefficient started to depend on the velocity, as predicted by existing ice friction models [8].

Schulson and Fortt [16] measured the friction coefficient of freshwater polycrystalline ice sliding slowly (10$^{-3}$ ~ 10$^{-8}$ m/s) upon itself at temperatures from 98 to 263 K under low normal stresses (≤ 98 kPa). The coefficient of kinetic friction of smooth surfaces varies from 0.15 to 0.76 and, at elevated temperatures (≥ 223 K), exhibits both velocity strengthening at lower velocities (10$^{-5}$ to 10$^{-4}$ m/s) and velocity weakening at higher velocities. At intermediate temperatures of 173 and 133 K, the kinetic coefficient appears to not exhibit significant dependence upon velocity. However, at the low temperature of 98 K the coefficient of kinetic friction exhibits moderate velocity strengthening at both the lowest and the highest velocities but velocity independence over the range of intermediate velocities.

Figure 2(b) shows that the friction coefficient of steel-pin on ice-disc in 10$^{-10}$ Pa vacuum depends linearly on temperature in the regime of solid bulk phase [6] but the coefficient (inset) exhibits insignificant temperature dependence in the bulk quasi-solid phase regime [7] under different conditions [8]. These temperature trends indicate the intrinsic behavior of ice at different structure phases.

3.2 Slipperiness of ice – Quasi-liquid skin

Researchers have heavily debated the seemingly simple question of why ice is slippery since 1850 when Faraday [1] firstly proposed that a liquid or a quasiliquid layer serves as the lubricant making ice slippery after his simple experiment: he pressed two cubes of ice against each other submerged in 0 °C water, and they fused together. Faraday argued that the liquid layers froze solid when they were no longer at the surface.

Intuition indicates that liquids are mobile and that their presence reduces friction between solids, which is why water spilled on a kitchen floor or rainwater on asphalt or concrete can create the same kinds of hazards for walkers and drivers that ice can. Therefore, in order to make that solid slippery a liquid must form on it that allows skates to slip. Therefore, Fraday's proposal is deemed true. Presumably, the liquid makes the surface slippery because liquids are mobile, whereas solid surfaces are relatively rigid. Asking why ice is slippery is thus roughly equivalent to asking how a liquid or quasiliquid layer can occur on the ice surface in the first place.

How is that thin layer of liquid water going to appear if ice's temperature is well under its melting point? Bob Rosenberg [17], an emeritus professor of chemistry at Lawrence University in Appleton, Wisconsin, featured in 2005 *Physics Today* on the history and progress on "why ice is slipper" in terms of pressure melting [2], frictional heating [4], and intrinsic quasiliquid forming or premelting [17].

## 3.3 Pressure melting

The conventional explanation, pressure melting, was suggested by James Thomson [2] in 1850 and lately experimentally approved by his brother, William Thomson, Later Lord Kevin [3], in 1850 as a consequence of the higher density of liquid water relative to ice. James Thomason [2] calculated that a pressure of 466 atmospheres would correspond to a melting pressure of -3.5 °C. Lord Kelvin [3] verified that result experimentally. However, he was not able to explain how hockey players and figure skaters were able to slide at temperatures below -3.5 °C at which temperature and below no pressure melting takes place. Skating is possible at very cold from around -30 °C, so how is it possible for skaters to skate at this very cold temperature? Their own weight would not be able to pressure the ice enough to drop the melting temperature of ice and create a thin layer of liquid water. The pressure-melting explanation also fails to explain why someone wearing flat-bottom shoes, with a much greater surface area that exerts even less pressure on ice can also slip on the ice.

The optimum temperature for figure skating is -5.5 °C and for hockey, -9 °C; figure skaters prefer slower, softer ice for their landings, whereas hockey players exploit the harder, faster ice. Indeed, skating is possible in climates as cold as -30 °C and skiing waxes are commercially available for such low temperatures. In his 1910 account of his last expedition to the South Pole, Robert Falcon Scott [17] tells of skiing easily at -30 °C though the snow surface is sand-like at -46 °C. But surprisingly, even with little evidence in its favor, pressure melting was dominant for more than a century and still remains as the dominant explanation of the slipperiness of ice in many text books.

## 3.4 Friction heating

Bowden and Hughes [4] proposed in 1939 the frictional heating mechanism. Friction is the force that generates heat whenever two objects slide against each other. If you rub your hands together, you can warm them up. When a skate moves on the surface of ice, the friction between the skate and the ice generates heat that melts the outermost layer of ice. Bowden and Hughes [4] suggested that the frictional heating dominates alternatively ice slippery. They did an experiment at a research station in Switzerland to maintain temperatures below -3°C using solid $CO_2$ and liquid air. Using surfaces of wood and metal, they measured the effects of static and kinetic friction on ice melting. They concluded that frictional heating was responsible for melting ice. Although frictional heating may answer why ice is slippery when moving, this theory does not explain why ice can be so slippery even for someone standing still on it.

## 3.5 Quasi-liquid skin of undercoordinated molecules

Michael Faraday [1] suggested that a film of water on ice would freeze when placed between two pieces of ice, but that the film would remain liquid on the surface of a single piece. However, he was not able to reason at the molecular level why the liquid layer forms without friction heating or pressure melting.

In 1949, Gurney [5] suggested that an intrinsic liquid film plays a role in the slipperiness of ice, which provided the state-of-the-art mechanism for lubricating ice and has been intensively studied in recent decades [6-10]. Gurney hypothesized that molecules, inherently unstable at the surface due to the lack of molecules above them (molecular undercoordination), migrate into the solid until the surface becomes stable, which prompts the formation of a liquid phase. If appreciable atomic migration takes place, the surface of a crystalline solid melts, like surface melting point depression happened to most normal substance [18], and the solid is covered with a thin liquid film under a tension force greater than that of the corresponding supercooled liquid. This tension force is numerically equal to the free energy of the surface. If such a solid is subsequently cooled to a temperature at which atomic migration effectively ceases, it will have frozen in its surface a tension force corresponding to thermal equilibrium at some higher temperature.

Since mid-1960s, a variety of experimental approaches, performed under various conditions, has been brought to bear on the premelting problem to determine the temperature range and thickness of the postulated quasiliquid layer. In 1969, Orem and Adamson [19] found that impurity adsorption promotes surface melting. Physical adsorption of simple hydrocarbon vapors on ice creates a liquid-like layer on the surface of ice. The adsorption of *n*-hexane on the surface of ice can form liquid-like layer at temperatures above -35 °C. These researchers interpreted their results as indicating that the onset of ice's surface premelting is at -35 °C. In the 1990s, chemistry Nobel laureate Mario Molina and coworkers [20] attributed the adsorption of hydrochloric acid on polar stratospheric clouds to the existence of a liquid-like layer on ice, which plays a role in the destruction of ozone.

### 3.5.1  NMR, XRD, proton and electron diffraction

Nuclear magnetic resonance (NMR)[21] provided evidence for liquid layer formation on ice: below the melting point there is a narrow absorption line, not the broad line one would expect from a periodic solid. Molecules at the surface between -20 °C and 0 °C rotate at a frequency five orders of magnitude greater than those in bulk ice and about 1/25 as fast as those in liquid water. The self-diffusion coefficient is two orders of magnitude larger than that in bulk ice. Using proton backscattering, Golecki and Jaccard [22] found in 1977 that surface vibrations of the oxygen atoms are roughly 3.3 times the amplitude of their bulk value, and estimated an amorphous layer 10 times thicker than what NMR measurements had estimated. But, unlike NMR, the proton backscattering measurements were made under high vacuum, a condition markedly different from the finite vapor pressures at which surface melting typically occur. Molecules perform differently with the ambient vapor pressure.

X-ray diffraction study [23] done in 1987 suggested that the intermolecular distance on the ice surface is slightly shorter than it is in liquid water but smaller than that of ice's bulk interior. In the mid-1990s Helmut Dosch and coworkers [24] found a liquid-like layer on the different crystallographic ice surfaces between -13.5 °C and 0 °C. The surface layer exhibits rotational disorder with intact long-range positional order well below the surface melting temperature. At the surface-melting temperature, a completely disordered layer exists on the surface above the rotationally disordered layer.

Experiments conducted in 1996 by Gabor Somorjai and coworker [25] also suggested the presence of quasiliquid layer when they probed the surface of thin layers of ice with low-energy electron diffraction, a technique that uses electrons to determine the surface structure of a crystal in the same way as x-ray diffraction reveals the crystal structure of a solid. By observing how electrons bounced off ice surface, they suggested that the rapidly vibrating oxygen ions actually make the surface of ice slippery. These "liquid-like" water molecules do not move from side to side - only up and down. If the atoms moved from side to side, the layer would actually become liquid, which is what happens when the temperature rises above 0 °C.

### 3.5.2  AFM friction

In 1998, using atomic force microscopy, Döppenschmidt and Butt [26], measured the thickness of the liquid-like layer on ice, in a temperature range above -35 °C. As illustrated in Figure 6, capillary contacting forces on the liquid surface prompted the cantilever tip of the atomic force microscopy (AFM) to jump into contact with the solid ice once it reached the much softer layer's level. The upper limit in thickness of the liquid-like layer varied from 70 nm at -0.7 °C, 32 nm at -1 °C, to 11 nm at -10 °C. Their results indicated that at about -33 °C surface melting starts. The temperature dependence of $d$ follows roughly $d \propto -\log \Delta T$, where $\Delta T$ is the difference between the melting temperature and the actual temperature. The addition of salt increased the thickness of the liquid-like layer. However, dragging the tip of an atomic force microscope across the surface of ice derived high friction of ice, which indicates that while the top layer of ice may be liquid, it is too thin to contribute much to slipperiness except near the melting temperature.

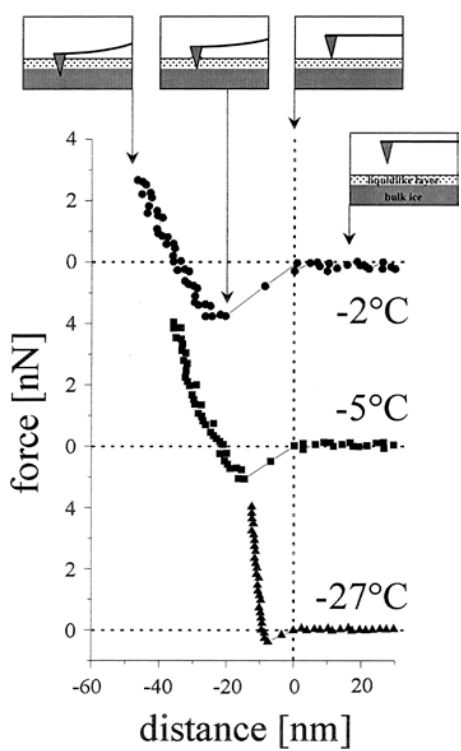

Figure 6. Approaching part of force curves measured at different temperatures. "Zero distance" was defined at the surface of the liquid-like layer. The assumed position of the tip is indicated schematically for the force curve taken at −2 °C.(Reprinted with permission from [26].)

3.5.3 X-ray reflection

However, Engemann and coworkers [27] studied in 2004 the interface between ice and solid silicon dioxide using x-ray reflectivity and calculated the thickness and density of the liquid layer between -25°C and 0°C, see

Figure 7. They found the density of the quasisolid skin varied from that of liquid water at its melting point to 1.16 g/cm$^3$ at -17°C, like a "high-density form of amorphous ice". The thickness of the quasiliquid layer follows the relationship,

$$L(T) = \alpha Ln \frac{17 \pm 3}{T_m - T} (nm)$$

Where $\alpha$ (0.84 ± 0.02 nm) is the decay factor. This experiment supports quasiliquid skin covering ice as the main cause of ice's slipperiness observed at temperature at -17 °C and above.

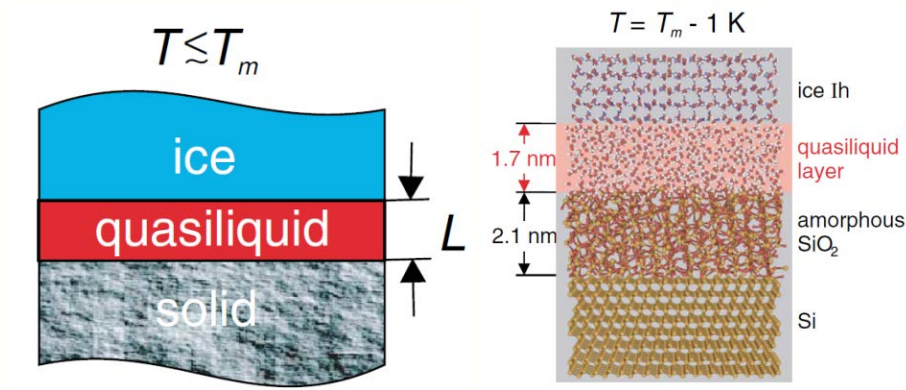

Figure 7. Quasiliquid skin forms between ice and amorphous SiO$_2$ at T ≥ T$_m$ – 17 K and its thickness increases with temperature. (Reprinted with permission from [27].)

3.6 A common supersolid skin covers both water and ice

From the perspective of O:H-O bond relaxation between undercoordinated water molecules, Sun and coworkers [28] proposed in 2009, and verified subsequently [9, 10] using quantum theory calculations and electron and phonon spectrometrics an elastic Coulomb-levitation mechanism for ice slippery as explained and detailed in next section.

4   Quantitative evidence

4.1 Ice skin segmental bond length and low mass density

Generally, bond order loss shortens and stiffens bonds between undercoordinated atoms by up to 12% for a flat skin of fcc geometry, which enhances the bond energy by 45% and depresses the atomic cohesive energy by 62% for a metal like gold and copper. The enhanced bond energy raises the skin elasticity by 67% and depressed cohesive energy lowers the local meting temperature by 63% [11]. However, for water and ice, molecular undercoordination shortens and stiffens the H-O bond and lengthens and softens the O:H nonbond from their standard bulk values of 1.0004 and 1.6946 Å at 4 °C and the ambient pressure (0.1 MPa)[29].

Figure 9 features the residual length spectra (RLS) for the MD-derived $d_x$ of ice. Subtracting the length spectrum calculated using the 360-molecular unit cell without skin from that with a skin resulted in the RLS (Figure 8). The RLS indicates that the $d_H$ contracts from the bulk value of about 1.00 to about 0.95 Å at the skin, while the $d_L$ elongates from about 1.68 to about 1.90 Å, with high fluctuation as a broad peak. This cooperative relaxation lengthens the O-O by 6.8% (=1-(0.95 +1.90)/(1.0+ 1.68)) and is associated with an 82% density ($\rho^{1/3} \propto d_{OO}$). The peak of $d_H = 0.93$ Å even corresponds to the undercoordinated H-O radicals, whose vibration frequency is around 3650 cm$^{-1}$ [10].

The skin O:H-O relaxation in fact lengthens the O-O distance $d_{O-O}$ and lowers the mass density $\rho$ [30-34]. According to the density-geometry-length correlation of molecules packed in water and ice [29], the measured $d_{O-O}$ of 2.965 Å [32] for liquid water gives rise to segmental lengths of $d_H = 0.8406$ Å and $d_L = 2.1126$ Å, which correspond to a 0.75 g·cm$^{-3}$ skin mass density [9]. In comparison, the MD derivatives in Figure 9 are 0.82 g·cm$^{-3}$ density compared to a density of 0.70 g·cm$^{-3}$ for a dimer. These values, 0.75 – 0.82 g·cm$^{-3}$, are much lower than 0.92 g·cm$^{-3}$ for bulk ice. MD calculations confirmed the length trend due to molecular undercoordination despite quantitative deviation from true situation.

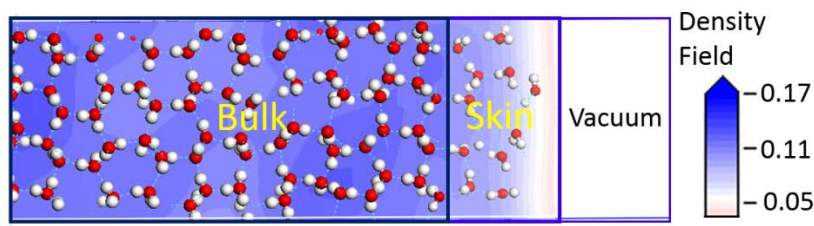

Figure 8. Schematic illustration of the water supercell with an insertion of a vacuum slab representing the supersolid skin of ice at 200 K. This comprised three regions, l. to r.: the bulk, the skin, and the vacuum. The skin contains undercoordinated molecules and free H-O radicals. The colors along the horizontal axis indicate the MD-

derived mass density field in the unit cell. This unit cell also applies to the shell of a nanobubble (reprinted with permission from [9]).

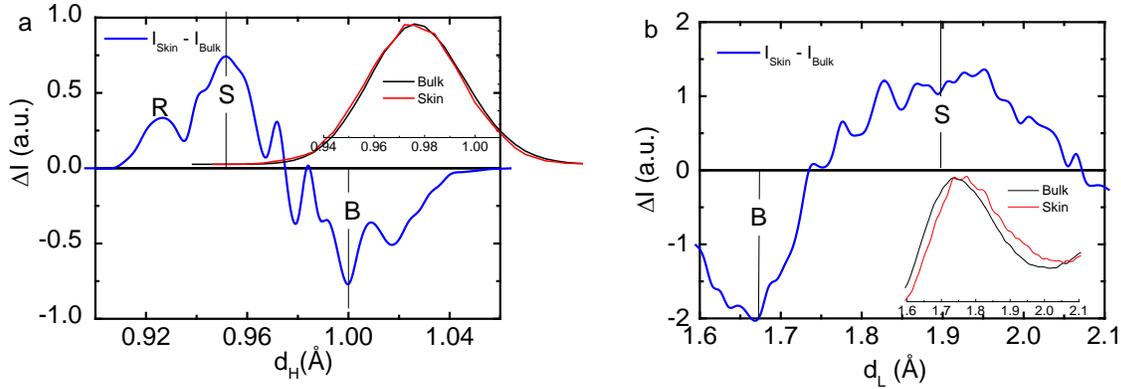

Figure 9. MD-derived RLS reveals that (a) $d_H$ contracts from the bulk value (B) of $\approx 1.00$ to $0.95$ Å for the skin (S) and to $0.93$ Å for the H-O free radicals (R), which is coupled with (b) $d_L$ elongation from the bulk value (B) of 1.68 to 1.90 Å, with high fluctuation. The insets show the raw spectra of the unit cell with skin (denoted 'skin') and without skin (denoted 'bulk'). (Reprinted with permission from [9].)

## 4.2 Identical $\omega_H$ for the skins of water and ice

Figure 10 (a, b) features the RPS for ice in comparison to (Figure 11) the measured $\omega_H$ RPS for both water and ice [35]. The valleys of the RPS represent the bulk feature, while peaks feature the skin attributes. A proper offset of the calculated RPS is necessary, as the MD code overestimates intra- and intermolecular interactions [7]. As expected, $\omega_L$ undergoes a redshift, while the $\omega_H$ undergoes a blueshift with three components. The $\omega_H$ blueshift results from the stiffening of the skin H-O bonds (S) and the free H-O radicals (R). The $\omega_L$ redshift arises from O-O repulsion and polarization. The polarization in turn screens and splits the intramolecular potential, which adds another $\omega_H$ peak (denoted P as polarization) with frequency being lower than that of the bulk valley (B), which was ever regarded as a second type of the O:H nonbond.

Most strikingly, the measured RPS in Figure 11 shows that the skins of water and ice share the same $\omega_H$ value of 3450 cm$^{-1}$, which indicates that the H-O bond in both skins is identical in length and energy, since $\omega_H \propto (E_H/d_H^2)^{1/2}$. The skin $\omega_L$ of ice may deviate from that of liquid water because of the extent of polarization, although experimental data is absent at this moment. Nevertheless, the skin $\omega_H$ stiffening agrees with the DFT-MD derivatives that the $\omega_H$ shifts from $\approx 3250$ cm$^{-1}$ at 7 Å depth to $\approx 3500$ cm$^{-1}$ of the 2 Å skin of liquid water [36]. Therefore, it is neither the case that an ice skin forms on water nor the case that a liquid skin covers ice. Rather, an identical supersolid skin covers both.

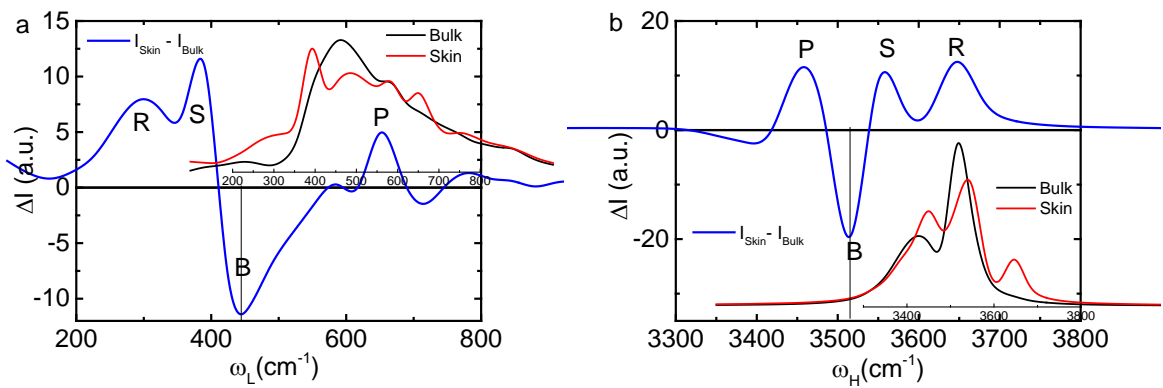

Figure 10. RPS of the MD-derived (a) $\omega_L$ and (b) $\omega_H$ of ice. The insets in (a) and (b) show the raw spectra of calculation. Calculations show that the $\omega_L$ undergoes a redshift, while the $\omega_H$, splitting into three, undergoes a blueshift. Features S corresponds to the skin H-O bond; R corresponds to the free H-O radicals; the P component arises from the screening and splitting of the crystal potential by the polarized nonbonding electrons. The skins of water and ice share the same $\omega_H$ of 3450 cm$^{-1}$. The peak intensity changes with the scattering from ice and water.

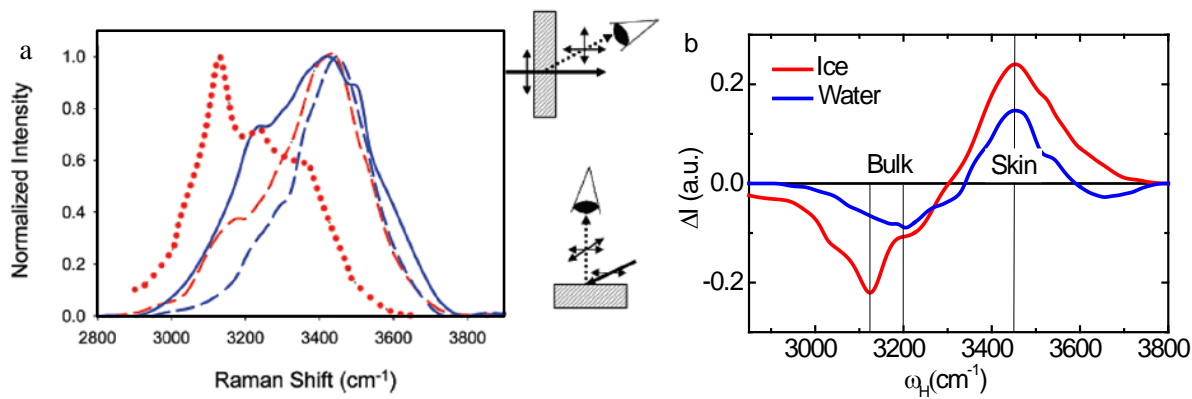

**Figure 11** (a) Raman $\omega_H$ spectroscopy of water (in blue, at 25°C) and ice (red, at -20 and -15°C) [35] collected at 87° (peaks toward higher frequency) and 0° with respect to the surface normal and water (side views). (b) The RPS of water and ice distills the skin peak from the bulk as valley contribution to the spectra [9].

### 4.3 Skin electron entrapment versus H-O bond energy

Table 1 features the DFT-derived Mulliken charge accumulation at the skin and in the bulk of water. O increases its net charge from the bulk value of -0.616 to -0.652 e for the skin. The net charge of a water molecule increases from 0.022 to -0.024 e correspondingly, which confirms the first round polarization of the electron lone pair by the entrapped O1s core electrons due to H-O bond contraction [9].

Table 1. DFT-derived charge localization at the skin and in the bulk of ice and derivatives (in bold) based on the referenced data using Eq. (1). Negative sign represents net electron gain.

|  | Skin | Bulk | $(H_2O)_1$ | O atom |
|---|---|---|---|---|
| $q_O$ | -0.652 | -0.616 | – | – |
| $q_H$ | 0.314 | 0.319 | – | – |
| Net $q$ of $H_2O$ | -0.024 | 0.022 | – | – |
| $E_{1s}$ (eV) [37-39] | 538.1 | 536.6 | 539.7 | **525.71** |
| $E_H$ (eV) | **4.52**/4.66 | 3.97 [29] | 5.10 [40] | – |
| $T_m$ (K) | **311**/320 | 273 | – | – |

The following formulates the skin H-O bond energy $E_H$(Skin) and the atomic O 1s energy $E_{1s}(0)$. Table 1 features the yields obtained with the known referenced data [18]:

$$\frac{\Delta E_{1s}(N)}{\Delta E_{1s}(\infty)} = \frac{E_{1s}(N) - E_{1s}(0)}{E_{1s}(\infty) - E_{1s}(0)} = \frac{E_H(N)}{E_H(\infty)} = \frac{T_C(N)}{T_C(\infty)} = \left(\frac{d_H}{d_{H0}}\right)^{-m}.$$

$$(1)$$

The $E_H$(Skin) = 3.97 × (538.1/536.6) = 4.52 eV is compatible with the value of 4.66 eV for breaking the H-O bond of $H_2O$ molecules deposited on a $TiO_2$ surface in less than a monolayer coverage using laser excitation [40]. The deviation $\Delta E_H$(Skin) = 0.14 eV (about 3%) arises mainly from molecular undercoordination in these two situations — one is the water skin and the other is the even less coordinated molecules on the $TiO_2$ surface, which indicates that interaction between water molecules and the hydrophobic $TiO_2$ surface is very weak with the presence of an 5 ~ 10 Å thick air gap in the hydrophobic contacts [41].

With the known values of $(d_H, E_H)_{Skin}$ = (0.84 Å, 4.52 eV) and $(d_H, E_H)_{Bulk}$ = (1.0 Å, 3.97 eV) and the $E_H(1)$ = 5.10 eV, the bond nature index is estimated as $m$ = 0.744 and the $d_H(1)$ = 0.714 Å of a monomer. The densely and locally entrapped core electrons of the undercoordinated water molecules polarize in a dual-process the nonbonding electrons.

## 4.4 Skin thermal stability

Generally, atomic undercoordination depresses the critical temperature for phase transition of many substances because of the drop of atomic cohesive energy, $T_C \propto zE_z$, where z is the atomic coordination number and $E_z$ is the cohesive energy per bond. The phase transition includes liquid-solid, liquid-vapor, ferromagnetic, ferroelectric, and superconductive transitions [11]. The skin melting temperature $T_{m,s}$ drops or rises depending the nature of the chemical bond, $T_{m,s} / T_{m,b} = z_s / z_b C_z^{-m}$, where m is the bond nature index and $C_z = 2\left\{1 + \exp\left[\left(z - 12\right)/\left(8z\right)\right]\right\}^{-1}$ is the contraction coefficient of bond between undercoordinated atoms. According to this BOLS notation, the skin $T_{m,s}$ is 40% and 62% of the bulk metal (m = 1) and Silicon (m = 4.88) as the effective atomic CN of the top layer is 4 contracts by 12%) and the bulk is 12 for an fcc structure standard [18].

However, for water molecules, the $T_C$ is proportional to either $E_H$ or the $E_L$ only, depending on the nature of phase change, because of the 'isolation' of the $H_2O$ molecule by its surrounding lone pairs. For instance, $E_L$ determines the $T_C$ for evaporation $T_V$, as this process dissociates the O:H nonbond. The $E_H$ dictates $T_m(\text{Skin})$ that is estimated based on the correlation between the $T_C(N)$ and the $\Delta E_{1s}(N)$ from Eq. (1):

$$\frac{T_C\left(Skin\right)}{T_C\left(\infty\right)} = \frac{T_m\left(Skin\right)}{273} = \frac{E_H\left(Skin\right)}{E_H\left(Bulk\right)} = \frac{4.59 \pm 0.07}{3.97},$$

which yields the skin melting temperatures in the range of 315 ± 5 K. It is therefore not surprising that water skin performs like ice or glue at room temperature (298 K) and that the monolayer water melts at about 325 K [42]. This observation rules out the premelting of ice skin.

## 4.5 Skin viscoelasticity

The polarization of molecules enhances the skin repulsion and viscoelasticity. The high viscoelasticity and the high density of skin dipoles are essential to the hydrophobicity and lubricity at contacts [43]. According to the BOLS-NEP notation [11], the local energy densification stiffens the skin and the densely and tightly entrapped bonding charges polarize nonbonding electrons in dual process to form anchored skin dipoles [28].

Table 2 features the MD-derived thickness-dependent $\gamma$, $\eta_s$ and $\eta_v$ of ice films. Reducing the number of molecular layers increases them all. The O:H-O cooperative relaxation and associated electron entrapment and polarization enhances the surface tension to reach the value of 73.6 mN/m for five layers, which approaches the measured value of 72 mN/m for water skin at 25°C. Generally, the viscosity of water reaches its maximum at a temperature around the $T_m$ [44].

Table 2. Thickness-dependent surface tension $\gamma$ and viscosity $\eta$.

| Number of layers | 15 | 8 | 5 |
|---|---|---|---|
| $\gamma$ (mN/m) | 31.5 | 55.2 | 73.6 |
| $\eta_s$ ($10^{-2}$mN·s/m²) | 0.007 | 0.012 | 0.019 |
| $\eta_v$($10^{-2}$mN·s/m²) | 0.027 | 0.029 | 0.032 |

The negative charge gain and the nonbonding electron polarization provide electrostatic repulsive forces lubricating ice.

4.6 Skin repulsion and hydrophobicity

Measurements, shown in Figure 12, in fact verified the presence of the repulsive forces between a hydrated mica substrate and the tungsten contacts at 24°C [45]. Such repulsive interactions appear at 20% – 45% relative humidity (RH). The repulsion corresponds to an elastic modulus of 6.7 GPa. Monolayer ice also forms on a graphite surface at 25% RH and 25°C [46]. These observations and the present numerical derivatives evidence the presence of the supersolidity with repulsive forces because of bonding charge densification, surface polarization and $T_m$ elevation.

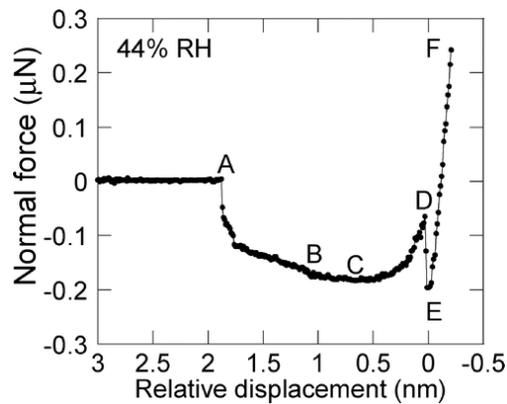

Figure 12. Normal force profiles between mica and tungsten tip at 44% RH. Point A is the initiation of water nucleation and condensation; B and C are the formation of a complete water bridge; D is the maximum attractive

force before the tip–substrate contact; E denotes the sudden drop of force; and F indicates the tip–substrate contact repulsive force. (Reprinted with permission from [45].)

## 4.7 Elastic Coulomb-levitation

It is convenient to adapt the concept of supersolidity from the superfluidity of solid $^4$He at mK temperatures. The skins of $^4$He fragments are highly elastic and frictionless with repulsion between them when set in motion [28]. The skins of water and ice form an extraordinary supersolid phase [9] that is elastic [35], hydrophobic [47, 48], polarized [49, 50] and thermally stable [42], with densely entrapped bonding electrons [37-39, 51] and ultra-low-density [32]. The fewer the molecular neighbors there are, the smaller the water molecule size is, the greater the molecular separation is, and therefore the greater the supersolidity will be.

### 4.7.1    Elastic Coulomb-levitation

According to the BOLS-NEP notation [52], molecular undercoordination shortens and stiffens the intramolecular H-O bond and meanwhile, lengthens and softens the intermolecular O:H-O bond because of the Coulomb repulsion between electron pairs on adjacent oxygen ions. The H-O will vibrate faster and the $(H_2O):(H_2O)$ slower at the skin. The dual polarization increases the local charge of O ions at the skin.

MD (Figure 9 and Figure 10) and DFT (Table 1) calculations confirmed as such. The O:H nonbond contracts from the bulk value of 1.0 to 0.95 Å for the skin and 0.93 for H-O radical and the H-O expands from 1.65 to 1.90 Å. The O:H phonon frequency shifts from the bulk value of 450 to 400 for the skin and to 300 cm$^{-1}$ for those close to free H-O radicals. The H-O phonon shifts from 3500 to 3550 and 3650 cm$^{-1}$ for the skin and H-O radicals, disregarding the artifact of the potential splitting and polarization effect.

Curvature conservation of an interatomic potential yields the relationship [11],

$$\mu(\omega x)^2 = \left( \frac{\partial^2 u(r)}{\partial r^2} \bigg|_{r=d} \right) x^2 = const\ ,$$

The O:H-O bond segmental vibration amplitudes of skin or radical vary from that of the bulk as,

$$\frac{x_{skin}}{x_{bulk}} = \frac{\omega_{bulk}}{\omega_{skin}} = \begin{cases} 450/400 = 9/8 & \text{(skin O:H)} \\ 450/300 = 3/2 & \text{(O:H radical)} \\ 3500/3550 = 70/71 & \text{(skin H-O)} \\ 3500/3650 = 70/73 & \text{(H-O radical)} \end{cases}$$

Therefore, the greater amplitudes and lower frequencies of the O:H soft springs and the undercoordination induced strong polarization are responsible for the slipperiness of ice, as illustrated in Figure 2b. The soft springs will deform easily when they are compressed and recover their original states once the sliding compression is relieved. If the compression force is too large, the O:H interaction will break, the coefficient will increase sharply.

### 4.7.2    High friction coefficient of ice on ice

As shown in Figure 1b, the kinetic friction coefficient of metal on ice rages from 0.01 to 0.1. Intuitively, friction coefficient (μ) of ice on ice could be even smaller. Actually it is the opposite. Firstly, Regelation takes place when two pieces of ice contact at a certain range of temperatures above -22 °C, if the pressure is sufficiently high, as observed by Faraday [1] who fused two pieces of ice in 0 °C water by slight compression. Secondly, water molecules tend to fuse as they recover their unoccupied neighbors, towards energetically favorable states. Finally, O:H phonon resonant coupling takes place when two pieces of ice are brought contact, as noted by Krim [14].

Kennedy and coworkers [53] reported that the friction coefficient of ice on ice varies with sliding velocity, 0.03 at 0.05 m/s and 0.58 at $5 \times 10^{-7}$ m/s within the temperature range of −3 °C and−40 °C under normal pressure of 0.007–1.0 MPa. Generally, μ decreased with increasing velocity and with increasing temperature, but was relatively insensitive to both pressure and grain size. The friction coefficients for freshwater and saltwater ice were almost indistinguishable at higher temperatures (−3°C and −10°C), but saline ice had lower friction at lower temperatures.

Schulson and Fortt [16] measured the friction coefficient of freshwater polycrystalline ice sliding slowly ($5 \times 10^{-8}$ to $1 \times 10^{-3}$ m s$^{-1}$) upon itself at temperatures from 98 to 263 K under low normal stresses (≤ 98 kPa). The coefficient of kinetic friction of smooth surfaces varies from $\mu_k = 0.15$ to 0.76 and, at elevated temperatures (≥ 223 K), exhibits both velocity strengthening at lower velocities ($<10^{-5}$ to $10^{-4}$ m s$^{-1}$) and velocity weakening at higher velocities. At intermediate temperatures of 173 and 133 K, the kinetic coefficient appears to not exhibit significant dependence upon velocity. However, at the low temperature of 98 K the coefficient of kinetic friction exhibits moderate velocity strengthening at both the lowest and the highest velocities but velocity independence over the range of intermediate velocities.

Therefore, it is not surprising why the friction coefficient of ice on ice is higher than the of ice contact by other materials.

### 4.7.3    Two-regime friction

Figure 1b shows two temperature regimes for the friction coefficient. At low temperature regime, the coefficient decreases with the rise of temperature, but at 250 K and above, the coefficient is insensitive to temperature. Why does this happen?

Figure 13 illustrates the O:H-O bond relaxation dynamics in the solid and in the quasisolid phases of bulk water. Retaining the same geometry of the bulk phase, the supersolid skin undergoes O:H elongation and H-O contraction, which is responsible for the slipperiness of ice. Inter-oxygen Coulomb repulsion and the segmental specific $\eta_x$ disparity of the O:H-O bond define that the H-O (x = H) bond and the O:H (x = L) nonbond relax simultaneously in the same direction but by different amounts [7]. The superposition of the specific heat curves yields two intersecting temperatures that define the boundaries of the solid/quasi-solid/liquid phases. The segment with a relatively lower specific heat follows the general rule of cooling contraction and the other segment relaxes performs oppositely.

In the quasisolid phase, $\eta_H/\eta_L < 1$, the H-O bond contracts at cooling, which lengthens the O:H nonbond and its vibrating amplitude, which enhances the slipperiness of ice. In the solid phase, $\eta_H/\eta_L > 1$, the O:H nonbond contracts at cooling and the H-O bond relaxes oppositely. The O:H nonbond shortening and its vibration amplitude reduction increases the friction coefficient of ice. This derivative clarifies why the friction coefficient show two temperature regimes – phase specific discrimination.

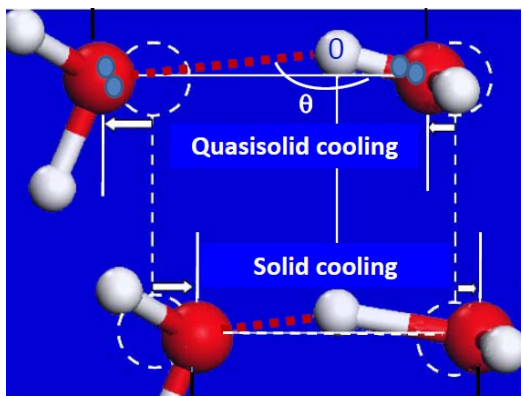

Figure 13. O:H-O bond relaxation in the solid (T < 258 K) and in the quasisolid (258 ≤ T ≤ 277 K) phases of bulk because of the segmental specific disparity [7]. O:H nonbond elongation and its vibration amplitude elevation lowers the friction coefficient.

### 4.7.4    AFM atomistic friction: sliding or scratching?

Generally, one talks about friction of an object sliding on ice, which gives lower friction coefficient. However, as shown in **Figure 6**, Atomic force microscopy in contacting mode derived high kinetic friction coefficient of 0.6 in the temperature range of -20 and – 40 °C, which is compatible to the static coefficient measured in macroscopic experiments [54]. The AFM tip scratching into the skin of several nanometers thick breaks the skin O:H nonbond with resistance of the high viscosity during scratching. The tip does not entertain the supersolid lubricating skin but the creep and viscosity resistance.

### 4.7.5    Curling physics (to be updated)

[55]

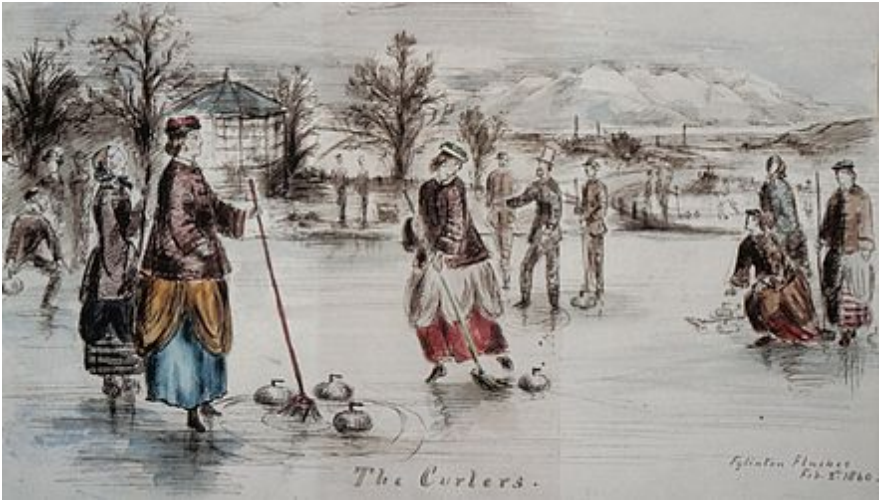

A curling match at Eglinton Castle, Ayrshire, Scotland in 1860. The Curling House is located to the left of the picture.

## 5    Insight extension

The supersolid skin is responsible for the slipperiness of ice and the hydrophobicity and toughness of water skin. This theory applies to the superfluidity of $^4$He [28], solid lubrication of oxides and nitrides, and water droplet flowing in carbon nanotubes [56]. Atomic undercoordination-induced local strain and the associated entrapment and polarization rationalize [4]He superfluidity - elastic and repulsive between locked dipoles at contacts. It is understandable now why the rate of the pressure-driven water flow through carbon nanotubes is orders higher in magnitude and faster than is predicted from conventional fluid-flow theory [57]. It is within expectation that the narrower the channel diameter is, the faster the flow of the fluid will be [56, 58], because of the curvature-enhanced supersolidity of the water droplet interacting with hydrophobic carbonnanotubes.

5.1 Supersolidity of [4]He crystals: elasticity and repulsivity

Helium is the noblest of elements: the interactions between even its own atoms are so weak that it solidifies only under intense pressure. If this pressure is reduced to below about 25 atmospheres at absolute zero, the quantum-mechanical fluctuations of the atoms' positions become so large that the solid melts, becoming a 'quantum liquid'. No crystalline solid is perfect — there are always some vacancies in the crystal lattice where atoms are missing — and in 1969 Alexander Andreev and Ilya Lifshitz [59] proposed that helium's large quantum fluctuations might, at zero temperature, stabilize a dilute gas of vacancies within the solid. Atoms of the prevalent isotope [4]He are bosons (they have zero spin), and so vacancies in solid [4]He can also be thought of as bosons. The vacancies can thus condense to form an exotic phase known as a Bose–Einstein condensate that suffuses the solid. This 'supersolid' phase would share some properties with a superfluid — namely, frictionless flow — but at the same time have a non-zero shear modulus, a defining characteristic of a solid. **Figure 15** illustrates the supersolid state of solid [4]He.

Supersolidity describes the coexistence of solid and superfluid properties in a quantum crystal. The phenomenon was discovered in 2004 by Eun-Seong Kim and Moses Chan [60, 61] when they measured the resonance period of a small cylindrical box oscillating around a torsion rod. The box contained solid [4]He, and below about 100 mK, the oscillation period decreased as if 1% of the helium mass had ceased moving with the box. The same method had been widely used for the detection of superfluidity in a liquid in the absence of viscosity, the liquid in the box remains at rest while the box walls move. At temperature below 200 mK, Helium 4 ([4]He) crystal is readily decoupled into fragments in a torsional oscillator to exhibit superfluidic nature – frictionless motion without viscosity [62-64]; meanwhile, the [4]He crystal fragments are stiffer than expected and hence react elastically to a shear stress applied [65]. The individual segment of the [4]He crystal would be thus both superelastic and superfluidic in motion - supersolidity.

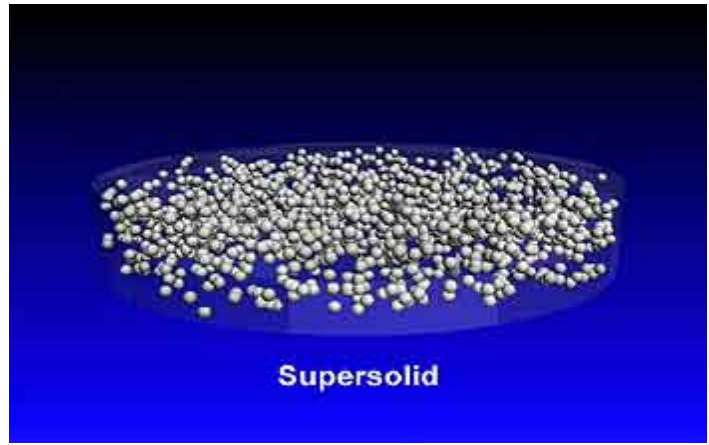

Figure 14. Supersolid of [4]He at 2 K temperatures or below [66]. The torsional oscillator is a disk filled with solid [4]He. To run the experiment, they hang the disk from a stiff rod and oscillate the disk back and forth. By measuring the frequency of oscillation, the scientists detect whether the solid [4]He behaves like a supersolid. An oscillating disk of normal matter, for example, behaves as expected: Because the atoms are rigidly linked, they rotate together. In an oscillating disk of supersolid matter, many of the atoms rotate, but some do not. Instead, those atoms slip through the lattice like a superfluid, with no friction whatsoever, and sit motionless. That reduces the mass of the disk, which allows it to oscillate faster. This animation has been exaggerated. In fact, the fraction of [4]He atoms that refuse to rotate is closer to only 1 percent. And the oscillation frequency Chan and Kim [60, 61] measured how many times the disk changes direction over a period of time is actually closer to 1000 times per second. The amplitude of the oscillation the distance the disk moves in either direction is not much bigger than the width of a single atom.

The 'supersolid' form of [4]He is stiffer, elastic, frictionless than the normal solid [67]. The superfluidity of [4]He solid is usually described in terms of Bose-Einstein condensation or quantum statistics in energy space. All particles occupy the lowest energy states simultaneously. A scenario in real space is infancy though the crystal defects have been recognized as the key to the supersolidity [68]. The superfluidity of [4]He solid is related to the quantum defects such as atomic vacancies of 1 nm size or around [69] and the supersolidity is related to structural disorder [70] such as dislocations, grain boundaries, or ill-crystallized regions where the under-coordinated atoms dominate. According to Pollet et al [69], inside a dislocation or a grain boundary, the local stress is anisotropic, which is sufficient to bring the vacancy energy to zero, so that the defect is invaded by vacancies that are mobile and superfluidic. Solid [4]He could contain a network of defects and if these defects are connected to each other, mass could flow from one side of the crystal to the other without friction. On the other hand, the disorder-induced stiffening could be the result of dislocations becoming pinned by isotopic impurities (i.e., [3]He atoms even at very small concentrations).

According to Anderson [68], the observations are conjectured to be describable in terms of a rarified Gross-Pitaevskii superfluid of vacancies, with a transition temperature of about 50 mK, whose density is locally enhanced by crystal imperfections. The observations can be affected by this density enhancement. Therefore, disorder and defects that could enhance the local density appear to play an important yet uncertain role in the supersolidity of $^4$He crystals [71].

The interatomic "bond" breaks easily for $^4$He crystals, which requires energy at the critical point of 4.2 K for liquid-vapor transition in the order of 1/3000 eV, much smaller than a typical van der Waals bond of 0.1 eV or around. The extremely weak interatomic interaction through charge sharing makes the $^4$He atoms or grains are sticking-less - more like hard spheres with closed electronic shell packing together. The sticking-less interaction between grains will lower the friction coefficient.

The understanding of slipperiness of ice and the BOLS-NEP notation at nanometer scaled contacts provide a mechanism for the superfluidity and supersolidity of $^4$He crystal. Repulsion between the "electric monopoles locked in the stiffened skins" of the small grains could help in solving this puzzle. Broken-bond-induced local strain and quantum entrapment leads to a densification of charge and energy in the skin of a few atomic layer thick. The densification of energy corresponds to the enhancement of the elasticity, which stiffens the solid skin allowing the $^4$He segment to react elastically to a shear stress. The repulsion between the densely entrapped electrons makes the motion frictionless. $^4$He crystals lack the nonbonding electrons because of the close atomic shells. Therefore, the broken bonds that serve as not only centers that initiate structure failure but also provide sites for pinning dislocations by charge and energy entrapment, which could be responsible for the superfluidity and supersolidity as observed. Its 'supersolid' behavior results just from atomic CN imperfection that changes the bulk properties of the crystal [72] - Atomic undercoordination induced local quantum entrapment and polarization.

The Coulomb repulsion between the "locked dipoles in the stiffened skins" of $^4$H fragments could help in understanding the puzzle of $^4$He crystal supersolidity in real space. The densification of energy corresponds to the enhancement of elasticity, which stiffens the solid skin allowing the $^4$He segment to react elastically to a shear stress; the repulsion between the charged surfaces makes the frictionless motion. The extremely weak interatomic interaction between the He atoms makes the $^4$He atoms or grains are non-sticky - more like hard spheres with close filled electronic shells. The sticky-less interaction between grains will lower the friction coefficient. Lattice contraction of the supersolid $^4$He segments is expected to happen, though this contraction is very small [63].

## 5.2 Superlubricity in dry sliding: atomistic friction

The ultralow-friction linear bearing of carbon nanotubes (CNTs) and the superlubricity at dry nanocontacts sliding in high vacuum [73, 74] are fascinating. As shown in Figure 16a, the velocity of the liquid water moving in the CNTs is inversely proportional to the diameter under constant pressure applied to the CNT end [57] which is beyond theory expectations. Transition electron microscopy revealed that the inner walls of a multi-walled CNT can slide back and forth with respect to the outer walls of the CNT, being free from wear for all cycles of motion (see Figure 16b) [75]. Surface energy calculations suggested that the force retracting the core nanotubes back into the outer tubes was 9 nN, being much lower than the van der Waals forces pulling the core nanotubes back into their sheath. The removal of the outer walls of the MWCNT corresponded to the highly localized dissipation at defect scattering sites, located primarily at the ends of the tube. The demonstration of ultralow friction between multi-walled CNT layers is a valuable confirmation that they will be useful mechanical components in molecular nanotechnology such as molecular bearing.

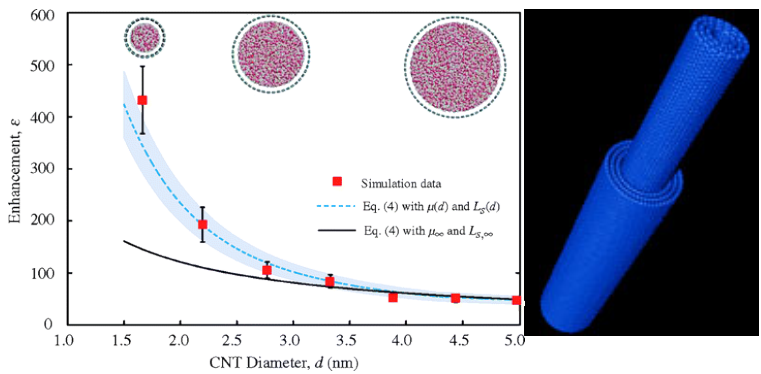

Figure 15 (a) Superfluidity of water droplet in CNTs of different diameters [57] and (b) ultralow-friction nanoscale linear bearing made of multi-walled CNT [75].

The occurrence of quantum friction is a kinetic process of energy dissipation ($E = f_r \cdot s$ with $f_r$ being the friction force and $s$ the sliding distance) due to the phonon (heat) and electron excitation (electron-hole pair production) during sliding [76]. A state of ultralow friction is reached when a sharp tip slides over a flat surface and the applied pressure is below a certain threshold, whose value is dependent on the surface potential sensed by the tip and the stiffness of the contacting materials [77-79].

A comparative study of hydrogen- and deuterium-terminated single-crystal diamond and silicon surfaces revealed that the hydrogenated surface (terminated with $H^+$) exhibited higher friction than the surface passivated with $^2H^+$.

The additional neutron in the $^2H^+$ should play a certain yet unclear role of significance because of the possible adsorbate size difference [76]. A remarkable dependence of the friction force on carrier type and concentration has been discovered by Park et al [80] on doped silicon substrates. An experiment of a biased conductive TiN coated tip of an atomic force microscope sliding on a Si substrate with patterned *p* and *n* stripes revealed that charge depletion or accumulation results in substantial differences in the friction force. A positive bias applied to the *p*-region causes a substantial increase of the friction force compared to the n region because of an accumulation of holes (+ charged) in the p region. No variation of friction force was resolvable between *n* and *p* regions under negative bias. Both observations [76, 80] indicate clearly that the positively charged ($H^+$) tip or substrate (electronic holes +) would induce high friction force.

The superlubricity phenomenon was explained using the classical Prandtl–Tomlinson (PT) model [81, 82] and its extensions, including thermal activation, temporal and spatial variations of the surface corrugation, and multiple-contact effects [77]. Observation suggests that the friction force depend linearly on the number of atoms that interact chemically across the contact [83]. According to the one-dimensional PT model, the slider atoms feel the periodic potential of the substrate surface atoms as they slide over them, experiencing a net force that is the sum of individual instantaneous friction force on each atom resulting from the gradient of the periodic potential.

The mechanism of interface electric repulsion also applies to the frictionless CNT linear bearing and the superlubricity of nanocontacts. In fact, the bond contraction happens to the CNT of limited number of walls. Bonds near the open ends contract even further. Densification of both the $\sigma$- and $\pi$-electrons takes place to all the walls; the repulsion between the densely packed and localized like charges will reduce the friction force substantially, while the electrostatic forces of the additionally densely charged CNT ends may provide force of retraction motion and oscillation. In the nanocontacts, the saturated potential barrier due to the skin charge trapping of the nanocontacts also provide a repulsion force between the contacts.

### 5.3 Solid lubrication

The key gradient of ice slippery is the presence of electron lone pairs and molecular undercoordination that softens the O:H nonbond with enlarged amplitude of vibration and high charge density due to dual polarization. Nitrides and oxide skins share the similarity of water and ice. N reacts with a solid skin preferring $C_{3v}$ symmetry, such as fcc(111) and hcp(0001) planes [84]. The N atom is located in a place between the top two layers and the lone pair is directed into the substrate. The surface is hence networked with the smaller $A^+$ and the saturate bonded $N^{3-}$ cores with densely packed electrons. Hence, the top skin layer is chemically inert as it is harder. Electrons in the saturated bond should be more stable compared with the otherwise unbonded electrons in the neutral host atoms.

The high intra-surface strength due to the ionic network could be responsible for the hardness of the top layer. On the other hand, the $N^{3-}$-$A^+$ network at the surface is connected to the substrate mainly through the nonbonding lone pair states. The nonbonding interaction is rather weak (~0.05 eV per bond) compared with the original metallic bonds (~1.0 eV per bond) or the intra-surface ionic bond (2~3 eV per bond). The weak lone-pair interaction is highly elastic within a critical load at which the weak interaction will break. Therefore, the enhanced intra-layer strength makes a nitride usually harder (~20 GPa), and the weakened inter-layer bonding makes the nitride highly elastic and self-lubricate. This mechanism also applies to graphite because of the weak C-axis interaction.

Nanoindentation profiles from TiCrN surface and sliding friction measurements from CN and TiN surfaces have confirmed the predicted high elasticity and high hardness at lower pressing load and the existence of the critical scratching load. As compared in **Figure 17**a, under 0.7 mN load of indentation, the elastic recoverability and hardness for a GaAlN film are higher than that of an amorphous carbon film [85]. The GaAlN surface is also much harder than the amorphous-C film under the lower indentation load. **Figure 17**b shows the profiles of pin-on-disk sliding friction test, which revealed the abrupt increase of the friction coefficient of nitride films when the load reaches critical value. For polycrystalline diamond thin films, no such abruption in friction coefficient is observed though the friction coefficient is generally higher than the nitride films. The absence of lone pairs in a-C film makes the film less elastic than a nitride film under the same pressing load. The abrupt change in the friction coefficient evidences the existence of critical load that breaks the nitride interlayer bonding – lone pair interaction. Therefore, the non-bonding interlayer interaction enhances the elasticity of nitride surfaces. Such high elasticity and high hardness by nature furnishes the nitride surfaces with self-lubricate for nano-tribological applications.

The mechanism of slipperiness of ice is analogous to the self-lubrication of metal nitride [85, 86] and oxide [87] skins with electron lone pairs coming into play. A 100% elastic recovery also presents to TiCrN, GaAlN and $\alpha$-$Al_2O_3$ skins under the critical frictional load (e.g., < 5 N for carbon nitride) at which the lone pair breaks.

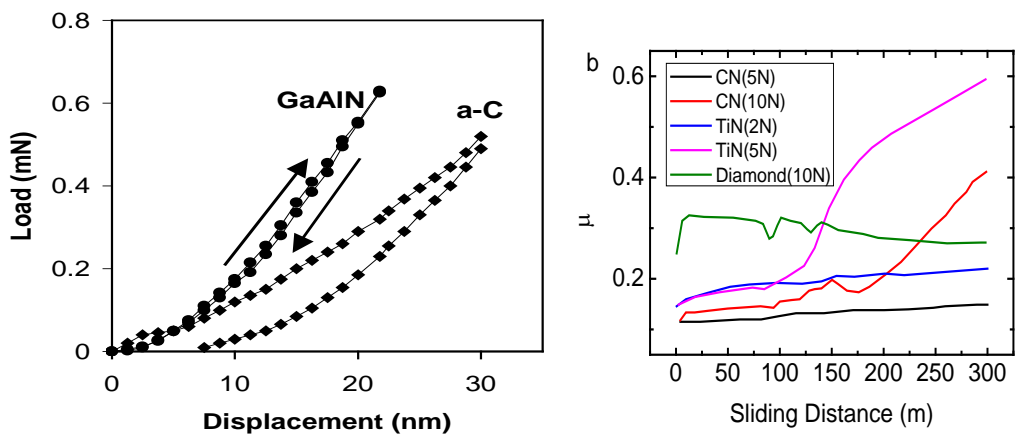

Figure 16. Solid lubricants. Comparison of (a) elasticity between GaAlN/Al$_2$O$_3$ films and amorphous carbon and (b) the pin-on-disc measurements of friction coefficient of nitrides under different loads in comparison to that of a diamond film. Lowering the operating temperature from the ambient (b) may reduce nitrides' friction coefficient to be compatible with ice. The abrupt increase of the coefficient indicates the presence of the critical load at which the lone pair nonbond breaks (Reprinted with permission from [85] and references therein.)

## 6    Summary

Molecular undercoordination-induced O:H-O bond relaxation and the associated nonbonding electron dual polarization clarify the skin supersolidity of ice. Agreement between numerical calculations and experimental observations verified the following:

1)    Undercoordination-induced O:H-O relaxation results in the supersolid phase that is elastic, hydrophobic, thermally more stable, and less dense, which dictates the unusual behaviour of water molecules at the boundary of the O:H-O networks or in the nanoscale droplet.

2)    The dual polarization makes ice skin hydrophobic, viscoelastic, and frictionless.

3)    Neither a liquid nor a quasiliquid skin forms on; rather, a common supersolid skin covers both water and ice. The supersolid skin causes slippery ice through the elastic Coulomb-levitation mechanism. The elastic, soft O:H nonbond springs of low frequency and high amplitude of vibration attached with pinned dipoles have high recoverability of reformation.

4)    These understanding extend to the superfluidity of $^4$He and the lubricity of water droplet flow in carbon nanotubes.